\documentstyle[12pt]{article}
\mathchardef\SGamma="7100

\begin{document}
\title{\vskip-1.7cm \bf  Cosmological constant problem and long-distance modifications
of Einstein theory}
\date{}
\author{A.O.Barvinsky}
\maketitle \hspace{-8mm} {\em Theory Department, Lebedev Physics
Institute, Leninsky Prospect 53, Moscow 119991, Russia}

\begin{abstract}
We construct the covariant nonlocal action for recently suggested
long-distance modifications of gravity theory motivated by the
cosmological constant and cosmological acceleration problems. This
construction is based on the special nonlocal form of the
Einstein-Hilbert action explicitly revealing the fact that this
action within the covariant curvature expansion begins with
curvature-squared terms.
\end{abstract}

New approach to the solution of the cosmological constant problem
consists in the assumption that, instead of adjusting the vacuum
energy of quantum matter to zero (or to small value of the
cosmological acceleration), one should modify the purely
gravitational sector of the theory in far infrared region. Matter
sources with wavelengths comparable with the horizon size of the
present Universe $L\sim 1/H_0\sim 10^{28}$ cm gravitate with the
long-distance gravitational constant $G_{LD}$ which is much
smaller than the conventional Planckian value $G_P$. Therefore,
the vacuum energy ${\cal E}$, $T_{\mu\nu}={\cal E}g_{\mu\nu}$, of
TeV or even Planckian scale (necessarily arising in all
conceivable models with spontaneously broken SUSY or in quantum
gravity) will not generate a catastrophically big spacetime
curvature incompatible with the tiny observable Hubble constant
$H_0^2\sim G_{LD}{\cal E}\ll G_P\,{\cal E}$. This mechanism is
drastically different from the old suggestions of supersymmetric
cancellation of ${\cal E}$ \cite{Weinberg}, because it relies on
the fact that the nearly homogeneous vacuum energy gravitates very
little, rather than it is itself very small.

This idea implies that the gravitational coupling constant should
be promoted to the level of the operator, $G_P\Rightarrow
G(\Box)$, which for sake of covariance can be regarded as a
function of the covariant d'Alem\-bertian
$\Box=g^{\alpha\beta}\nabla_\alpha\nabla_\beta$, interpolating
between the Planck scale of the gravitational coupling constant
$G_P=1/16\pi M^2_P$ for local matter sources of size $\ll L$ and
the long distance gravitational constant $G_{LD}=G(0)$ with which
the sources nearly homogeneous at the horizon scale are
gravitating \cite{AHDDG}. The modified equations of motion were
suggested to have the form of Einstein equations
    \begin{eqnarray}
    M^2(\Box)
    \left(R_{\mu\nu}-\frac12
    g_{\mu\nu}R\right)=\frac12\,T_{\mu\nu}  \label{1.1}
    \end{eqnarray}
with "nonlocal" inverse gravitational constant or Planck mass,
    \begin{eqnarray}
    \frac1{16\pi G(\Box)}\equiv M^2(\Box)=M_P^2\,
    \big(1+{\cal F}(L^2\Box\,)\big),
    \end{eqnarray}
being some function of the dimensionless combination of $\Box$ and
the additional scale $L$ -- the length at which infrared
modification becomes important. If the function of $z=L^2\Box$
satisfies the conditions, ${\cal F}(z)\to 0$ at $z\gg 1$, and
${\cal F}(z)\to {\cal F}(0)\gg 1$ at $z\to 0$, then the
long-distance modification is inessential for processes varying in
spacetime faster than $1/L$ and is large for slower phenomena at
wavelengthes $\sim L$ and longer.

One difficulty with this construction is that for any nontrivial
form factor ${\cal F}(L^2\Box\,)$ the left hand side of
(\ref{1.1}) does not satisfy the Bianchi identity and, therefore,
cannot be generated by generally covariant action. Obviously, this
makes the situation unsatisfactory because of a missing off-shell
extension of the theory, problems with its quantization, etc. Here
we suggest to circumvent this problem by resorting to the {\em
weak field} approximation, which is certainly valid in the
infrared regime. This means that Eq. (\ref{1.1}) should be
understood only as a first term of the perturbation expansion in
powers of the curvature. Its left hand side should be modified by
higher than linear terms in the curvature, and the modified
nonlocal action $S_{NL}[\,g\,]$ should be found from the
variational equation
    \begin{eqnarray}
    \frac{\delta S_{NL}[\,g\,]}{\delta g_{\mu\nu}(x)}=
    M_P^2\,g^{1/2}\Big(1+{\cal F}(L^2\Box\,)\Big)
    \left(R^{\mu\nu}-\frac12
    g^{\mu\nu}R\right)+{\rm O}\,[\,R_{\mu\nu}^2\,]. \label{1.3}
    \end{eqnarray}

Flexibility in higher orders of the curvature allows one to
guarantee the integrability of this equation and to construct the
nonlocal action as a generally covariant (but nonlocal) curvature
expansion. Here we explicitly present this construction along the
lines of covariant curvature expansion of \cite{CPT}. As a
starting point we consider a special nonlocal form of the
Einstein-Hilbert action revealing its basic property -- the
absence of a linear in metric perturbation part (on flat-space
background). Then we introduce a needed long-distance modification
by a simple replacement of the nonlocal form factor in the
curvature-squared term of the obtained action \cite{nonloc}. The
paper is accomplished by a discussion of the nature of
nonlocalities in quantum-gravitational and brane-induced models of
\cite{DGP}. In particular, the fact that curvature expansion for
the action begins with the quadratic order is revisited from the
viewpoint of the running gravitational coupling constant and the
issue of acausality of nonlocal effective equations, raised in
\cite{AHDDG}, is reconsidered.

The Einstein action in the Euclidean asymptotically-flat spacetime
    \begin{eqnarray}
    S_E[\,g\,]=-M_P^2
    \int dx\,g^{1/2}\,R(\,g\,)+
    M_P^2\int_{|x|\to\infty} d\sigma^\mu\,
    \big(\partial^\nu
    h_{\mu\nu}-\partial_\mu h\Big).    \label{2.1}
    \end{eqnarray}
includes the bulk integral of the scalar curvature and the
Gibbons-Hawking surface integral over spacetime infinity,
$|x|\to\infty$. The latter is usually called the Gibbons-Hawking
action which in the covariant form contains the trace of the
extrinsic curvature of the boundary (with an appropriate
subtraction of the flat space background contribution). This
surface term guarantees the consistency of the variational problem
for this action which yields as a metric variational derivative
the Einstein tensor.

The action (\ref{2.1}) is explicitly linear in the curvature, but
this linearity is misleading, because its variational derivative
--- the Einstein tensor --- is also linear in the curvature.
Therefore, it is at least linear in metric perturbation on
flat-space background, $R_{\mu\nu}\sim h_{\mu\nu}$, and the
perturbation theory for $S_E[\,g\,]$ should start with the
quadratic order, ${\rm O}\,[\,h_{\mu\nu}^2\,]\sim {\rm
O}\,[\,R_{\mu\nu}^2\,]$. This is a well known fact from the theory
of free massless spin-2 field. Our goal is to make this
$h_{\mu\nu}$-expansion manifestly covariant by converting it to
the covariant {\it curvature} expansion. A systematic way to do
this is to use the technique of covariant perturbation theory of
\cite{CPT}, which begins with the derivation of the expression for
the metric perturbation in terms of the curvature.

Expand the Ricci curvature in $h_{\mu\nu}$ on flat-space
background
    \begin{eqnarray}
    &&R_{\mu\nu}=-\frac12\,\Box\,h_{\mu\nu}+\frac12\,\Big(
    \nabla_\mu X_\nu+
    \nabla_\nu X_\mu\Big)
    +{\rm O}\,[\,h_{\mu\nu}^2\,],         \label{2.3}
    \end{eqnarray}
$X_\mu\equiv \nabla^\lambda h_{\mu\lambda}-\frac12\,\nabla_\mu h$,
and solve it by iterations as a nonlocal expansion in powers of
the curvature. This expansion starts with the following terms
    \begin{eqnarray}
    h_{\mu\nu}=-\frac2{\Box}R_{\mu\nu}
    +\nabla_\mu f_\nu+\nabla_\nu f_\mu
    +{\rm O}\,[\,R_{\mu\nu}^2\,].        \label{2.4}
    \end{eqnarray}
Here $1/\Box$ denotes the action of the Green's function of the
{\em covariant metric-dependent} d'Alembertian on the space of
symmetric second-rank tensors with zero boundary conditions at
infinity and the term $\nabla_\mu f_\nu+\nabla_\nu f_\mu$ in
(\ref{2.4}) reflects the gauge ambiguity in this solution (it
originates from the harmonic-gauge terms in the right-hand side of
(\ref{2.3})).

Now restrict ourselves with the approximation quadratic in
$R_{\mu\nu}$ (or equivalently $h_{\mu\nu}$) and integrate the
variational equation for $S_E[\,g\,]$. Since the variational
derivative is at least linear in $h_{\mu\nu}$, $\delta S_E/\delta
g_{\mu\nu}\sim h_{\alpha\beta}$, the quadratic part of the action
in view of this equation is given by the integral
    \begin{eqnarray}
    S_E[\,g\,]=\frac12\int dx \,h_{\mu\nu}(x)
    \frac{\delta S_E[\,g\,]}{\delta g_{\mu\nu}(x)}
    +{\rm O}\,[\,R_{\mu\nu}^3\,].           \label{2.5}
    \end{eqnarray}
Substituting the Einstein tensor for $\delta S_E/\delta
g_{\mu\nu}$ and (\ref{2.4}) for $h_{\mu\nu}$ and integrating by
parts one finds that the contribution of the gauge parameters
$f_\mu$ vanishes in view of the Bianchi identity for the Einstein
tensor, and the final result reads
    \begin{eqnarray}
    S_E[\,g\,]=
    M_P^2\int dx\,g^{1/2}\,\left\{\,
    -\Big(R^{\mu\nu}
    -\frac12\,g^{\mu\nu}R\Big)\,
    \frac1{\Box}R_{\mu\nu}
    +{\rm O}\,[\,R_{\mu\nu}^3\,]\,\right\}.  \label{2.6}
    \end{eqnarray}
This is the covariant {\em nonlocal} form of the {\em local}
Einstein action \cite{brane,nlbwa,nonloc}. This nonlocal
incarnation of (\ref{2.1}) explicitly features the absence of the
linear in curvature term, which can be clarified by the
subtraction effect of the Gibbons-Hawking term.

In asymptotically-flat (Euclidean) spacetime with the asymptotic
behavior of the metric
    $g_{\mu\nu}=\delta_{\mu\nu}+h_{\mu\nu}$,
    $h_{\mu\nu}={\rm O}\,\left(1/|x|^{d-2}\right)$, $|x|\to\infty$,
the Gibbons-Hawking term in Cartesian coordinates can be
transformed to the bulk integral of the integrand
$\partial^\mu\big(\partial^\nu h_{\mu\nu}-\partial_\mu h\Big)$ --
the linear in $h_{\mu\nu}$ part of the scalar curvature. Similarly
to the above procedure this integral can be covariantly expanded
in powers of the curvature. Up to quadratic terms inclusive this
expansion reads \cite{nonloc}
    \begin{eqnarray}
    \int_\infty\! d\sigma^\mu
    \big(\partial^\nu
    h_{\mu\nu}-\partial_\mu h\Big)=
    \int dx\,g^{1/2}\left\{R
    -\Big(R^{\mu\nu}
    -\frac12\,g^{\mu\nu}R\Big)\frac1{\Box}R_{\mu\nu}
    +...\right\}.     \label{2.8}
    \end{eqnarray}
As we see, when substituting to (\ref{2.1}) its linear term
cancels the Ricci scalar part and the quadratic terms reproduce
those of (\ref{2.6}). Obviously, this type of expansion can be
extended to arbitrary order in curvature.

%\section{Long-distance modification of the Einstein action}
%\hspace{\parindent}
Long distance modification of the Einstein action that would
generate (\ref{1.3}) as the left-hand side of the gravitational
equations of motion now can be simply obtained from the nonlocal
form of the Einstein action (\ref{2.6}). It is just enough to make
the following replacement in the quadratic part of (\ref{2.6}),
$1/\Box\rightarrow(1+{\cal F}(L^2\Box))/\Box$. Indeed, the
variation of the Ricci tensor here and integration by parts
"cancel" $\Box$ in the denominator. All commutators of covariant
derivatives with the $\Box$ in ${\cal F}(L^2\Box)$ give rise to
the curvature-squared order which is beyond our control. This
recovers the Einstein tensor term of (\ref{1.3}) with the needed
"nonlocal" Planckian mass $M_P^2\Big(1+{\cal F}(L^2\Box\,)\Big)$.

Thus, the long-distance modification in question takes the form
    \begin{eqnarray}
    S_{NL}[\,g\,]=-\int dx\,g^{1/2}\left\{
    \Big(R^{\mu\nu}
    -\frac12 g^{\mu\nu}R\Big)
    \frac{M^2(\Box)}{\Box}\,
    R_{\mu\nu}
    +{\rm O}[R_{\mu\nu}^3]\right\}.  \label{3.2}
    \end{eqnarray}
It is manifestly generally covariant, and its variational
derivative (the left hand side of the modified Einstein equations)
exactly satisfies the Bianchi identity and does not suffer from
the concerns of \cite{AHDDG}. This action is not unique though,
because it is defined by a given form factor ${\cal F}(L^2\Box\,)$
only in quadratic order, while we do not have good principles to
fix its higher-order terms thus far.

%\section{Discussion: running coupling constants and nonlocality
%vs acausality}
%\hspace{\parindent}
One of the main mechanisms for nonlocalities of the above type is
the contribution of graviton and matter loops to the quantum
effective action. In quantum theory the concept of a nonlocal form
factor replacing a coupling constant is not new. In fact this
concept underlies the notion of the running coupling constants and
sheds new light on the cosmological constant problem also from the
viewpoint of the renormalization theory. For simplicity, consider
QED or Yang-Mills theory in the quadratic order in gauge field
strength $F_{\mu\nu}^2$. The transition from classical to quantum
effective action, $S\to S_{\rm eff}$, boils down to the
replacement of the local invariant by
    \begin{eqnarray}
    g^{-2}\int dx\,F_{\mu\nu}^2\to
    \int dx\,F_{\mu\nu}\,g_{\rm eff}^{-2}(-\Box)\,
    F^{\mu\nu}.
    \end{eqnarray}
Here the effective coupling constant $g_{\rm eff}^{-2}(-\Box)$ is
a nonlocal form factor which can be obtained from the
corresponding solution of renormalization-group equation.
Obviously, it plays the role of ${\cal F}(L^2\Box)$ above.

This concept, however, fails when applied to the gravitational
theory in the sector of the cosmological and Einstein-Hilbert
terms\footnote{When applied to formally renormalizable (albeit
non-unitary) curvature-squared gravitational models
\cite{Tseytlin}.}. Indeed, naive replacement of ultralocal
cosmological and gravitational coupling constants by nonlocal form
factors,
    \begin{eqnarray}
    \int dx\,g^{1/2}\left(\,\Lambda-M_P^2\,R\,\right)\to
    \int dx\,g^{1/2}\left(\,\Lambda(\Box)-M_P^2(\Box)
    \,R\,\right),
    \end{eqnarray}
is meaningless because the action of the covariant d'Alembertian
on the right hand side always picks up its zero mode, and both
form factors reduce to their numerical values in far infrared,
$\Lambda(0)$, $M_P^2(0)$. Therefore, even if one has solutions of
renormalization group equations for $\Lambda$ and $M_P^2$, like
those of \cite{Tseytlin}, one cannot automatically recover the
corresponding pieces of effective action or the corresponding
nonlocal correlation functions.

The construction above suggests that the running coupling constant
``delocalization'' of $M_P^2$ should be done in the (already
nonlocal) representation of the Einstein action (\ref{2.6}). As
its curvature expansion begins with the quadratic order, one can
insert the nonlocal form factor $M_P^2(\Box)$ {\em between two
curvatures} so that no integration by parts would result in its
degeneration to a trivial constant. It would be interesting to see
how a similar mechanism works for the nonlocal cosmological
``constant'' $\Lambda(\Box)$.\footnote{One should expect that a
quadratic action for the cosmological term would read as
$\Lambda\int dx\,g^{1/2}R^{\mu\nu}(1/\Box^2)R_{\mu\nu}$. This
structure (also suggested in \cite{shap} and discussed within the
renormalization group theory) appears in two-brane models
\cite{nlbwa} and expected as a covariant completion of the mass
term in models of massive gravitons \cite{completion} and
discussions of the van Damm-Veltman-Zakharov discontinuity
\cite{vDVZ}.} The mechanisms of its generation due to infrared
asymptotics of the effective action, or late-time asymptotics of
the corresponding heat kernel, are discussed in \cite{nneag}.

Nonlocalities of the type (\ref{3.2}) also arise in a certain
class of braneworld models \cite{GRS,DGP}. They cannot appear in
models of the Randall-Sundrum type with strictly localized zero
modes, because in these models nontrivial form factors basically
arise in the transverse-traceless sector of the action (as kernels
of nonlocal quadratic forms in {\em Weyl} tensor \cite{nlbwa}). In
contrast to these models, the nonlocal part of (\ref{3.2}) is not
quadratic in the Weyl tensor, $\int
dx\,g^{1/2}W^2_{\mu\nu\alpha\beta}\sim\int
dx\,g^{1/2}(R_{\mu\nu}^2-\frac13 R^2)$ (with the insertion of a
nonlocal form factor between the curvatures). Rather, (\ref{3.2})
includes the structure $\int dx\,g^{1/2}(R_{\mu\nu}^2-\frac12
R^2)$ which contains the {\em conformal} sector. It is this sector
which is responsible for the potential resolution of the
cosmological constant problem. It becomes dynamical in models with
metastable graviton like the Gregory-Rubakov-Sibiryakov model
\cite{GRS} or brane induced gravity models of the
Dvali-Gabadaze-Porrati (DGP) type \cite{DGP}. In particular, for
the (4+1)-dimensional DGP model with the bulk $G_{AB}(X)$ and
induced on the brane $g_{\mu\nu}(x)$ metrics
    \begin{eqnarray}
    S_{DGP}[\,G\,]=M^3\int d^5X\,G^{1/2}\;\vphantom{L}^5\!
    R(G_{AB})+M_P^2\int d^4x\,g^{1/2}\;\vphantom{L}^4\!
    R(g_{\mu\nu})
    \end{eqnarray}
(we disregard here relevant Gibbons-Hawking terms) the effective
braneworld action takes the form (\ref{3.2}) with the form factor
${\cal F}(L^2\Box)$ which is singular at $\Box\to 0$
\cite{DGP,DefDG}
    \begin{eqnarray}
    {\cal F}(L^2\Box)=
    \frac1{L\sqrt{-\Box}},\,\,\,\,\,
    L=\frac{M_P^2}{M^3},              \label{4.1}
    \end{eqnarray}
where $M\sim 10^{-21}M_P\sim 100$ MeV is a mass scale of the bulk
gravity as opposed to the Planckian scale of the Einstein term on
the brane $M_P\sim 10^{19}$ GeV. This model is interesting,
because its approximate cosmological FRW equation of motion,
    \begin{eqnarray}
    H^2-\frac{H}L=\frac{\rho}{6M_P^2}\, , \label{4.1a}
    \end{eqnarray}
admits the self-accelerating regime with $H=1/L$ at late stages of
evolution with matter density $\rho\to 0$ \cite{DefDG}.
Unfortunately, this model can hardly serve as a consistent
candidate for the description of our Universe because of the
presence of the sufficiently low strong-coupling scale
$(M_P/L^2)^{1/3}\sim 10^{-8} \mbox{cm}^{-1}$, which invalidates
its applications to phenomena already at distances smaller than
$1000$ km \cite{scale} (see, however, Ref. \cite{reg} advocating
that this difficulty can be circumvented by special type of
regularization).

Form factors like (\ref{4.1}) are unambiguously defined only in
the Euclidean space with the negative semi-definite d'Alembertian
$\Box$. This raises the problem of their continuation to the
Lorentzian spacetime where the issues of causality and unitarity
become important. The principles of this continuation depend on
the physical origin of nonlocality in ${\cal F}(L^2\Box)$. In
particular, scattering problems for in-out matrix elements,
$\big<\,out\,|\,\hat\varphi\,|\,in\,\big>$, in spacetime with
asymptotically-flat past and future imply a usual Wick rotation.
The problem for in-in mean value of the quantum field,
$\phi=\big<\,in\,|\,\hat\varphi\,|\,in\,\big>$, is more
complicated and incorporates the Schwinger-Keldysh diagrammatic
technique \cite{SchKel}. In this technique the effective equations
for $\phi$ cannot be obtained as variational derivatives of some
one-field action functional. However, there exists a special case
of the initial quantum state --- the {\em Poincare-invariant
in-vacuum} in asymptotic past, $|\,in\,\big>=|\,in,vac\,\big>$.
Effective equations for $\phi$ in this vacuum can be obtained by
the following procedure \cite{CPT}. Calculate the {\em Euclidean}
effective action in asymptotically-flat spacetime, take its
variational derivative containing the nonlocal form factors which
are uniquely specified by zero boundary conditions at Euclidean
infinity. Then formally go over to the Lorentzian spacetime
signature with the {\em retardation} prescription for all nonlocal
form factors. These retarded boundary conditions uniquely specify
the nonlocal effective equations and guarantee their causality.
This procedure was proven in \cite{CPT} and also put forward in a
recent paper \cite{Woodard} as the basis of the covariant nonlocal
model of MOND theory.

This procedure justifies the Euclidean setup used above and
suggests that in this setting no contradiction arises between the
nonlocal nature of the {\em Euclidean} action and {\em causal}
nature of nonlocal equations of motion in Lorentzian spacetime. In
this respect the situation is essentially different from the
assumptions of \cite{AHDDG} where acausality of equations of
motion is necessarily attributed to the nonlocal action. No such
assumptions are needed in effective equations for expectation
values which are fundamentally causal despite their nonlocality.
These equations have interesting applications in quantum
gravitational context and, in particular, show the phenomenon of
the cosmological acceleration due to infrared back-reaction
mechanisms \cite{W1}.

The situation with brane induced nonlocalities and their causality
status is more questionable and conceptually open. For example,
the branch point of the square root in the nonlocal form factor
(\ref{4.1}) is apparently related to different branches of
cosmological solutions of (\ref{4.1a}) including the scenario of
cosmological acceleration \cite{DefDG}. Therefore, in contrast to
tentative models of \cite{AHDDG} with finite ${\cal F}(0)\gg 1$,
which only interpolate between two Einstein theories with
different gravitational constants $G_{LD}\sim G_P/{{\cal F}(0)}\ll
G_P$, the DGP model is anticipated to suggest the mechanism of the
cosmological acceleration. This implies the replacement of the
asymptotically-flat spacetime by the asymptotically-deSitter one.
For small values of asymptotic curvature (as is the case of the
observable horizon scale $H_0^2/M_P^2\sim 10^{-120}$) the
curvature expansion used above seems plausible, although the
effect of the asymptotic curvature might be in essence
nonperturbative. Therefore, the above construction might have to
be modified accordingly. In particular, the expansion in powers of
the curvature should be replaced by the expansion in powers of its
deviation from the asymptotic value $R_{\mu\nu}-\frac1d
g_{\mu\nu}R_\infty$ ($d$ is a spacetime dimensionality). This
would introduce in the formalism as a free parameter the value of
the curvature in far future, $R_\infty\sim H_0^2$, reflecting the
measure of acausality in the model. The resulting modifications in
the above construction are currently under study and will be
presented elsewhere.
\\
%\section*{Acknowledgements}

The author would like to thank G.Gabadadze, V.Mukhanov,
S.Solodukhin and R.Woodard for helpful stimulating discussions.
This work was supported by the Russian Foundation for Basic
Research under the grant No 02-02-17054 and the LSS grant No
1578.2003.2.

\end{document}